# Statistical features of edge turbulence in RFX-mod from Gas Puffing Imaging

F. Sattin[#,1], P. Scarin[1], M. Agostini[1,2], R. Cavazzana[1], G. Serianni[1], M. Spolaore[1], N. Vianello[1]

[1]Consorzio RFX, Euratom-ENEA Association,
Corso Stati Uniti 4, 35127 Padova, Italy
[2]Dipartimento di Fisica, Università di Padova

**Abstract**

Plasma density fluctuations in the edge plasma of the RFX-mod device are measured through the Gas Puffing Imaging Diagnostics. Statistical features of the signal are quantified in terms of the Probability Distribution Function (PDF), and computed for several kinds of discharges. The PDFs from discharges without particular control methods are found to be adequately described by a Gamma function, consistently with the recent results by Graves *et al* [J.P. Graves, *et al*, Plasma Phys. Control. Fusion **47**, L1 (2005)]. On the other hand, pulses with external methods for plasma control feature modified PDFs. A first empirical analysis suggests that they may be interpolated through a linear combination of simple functions. An inspection of the literature shows that this kind of PDFs is common to other devices as well, and has been suggested to be due to the simultaneous presence of different mechanisms driving respectively coherent bursts and gaussian background turbulence. An attempt is made to relate differences in the PDFs to plasma conditions such as the local shift of the plasma column. A simple phenomenological model to interpret the nature of the PDF and assign a meaning to its parameters is also developed.

[#] Corresponding author. E-mail address: fabio.sattin@igi.cnr.it

# 1. Introduction

The physics of plasma transport in the edge region of fusion devices is of considerable interest for its implications in plasma-wall interactions and, ultimately, confinement performances of the devices themselves. One relevant fact that has attracted attention of researchers for several years is the observation that particle edge transport is intermittent, in the form of much-larger-than-average events occurring sporadically but more often than predicted from a Gaussian statistical distribution. Since these bursts, although rare, do contribute for a relevant fraction of the total flux, an understanding of the physics lying behind the generation and dynamics of the intermittent events is deemed necessary. Nowadays, it is commonly recognized that bursts are associated to coherent plasma structures moving within the plasma. A remarkable feature is that many statistical properties of these objects appear to be universal, i.e., common to all plasma devices, regardless of their magnetic (Tokamaks, Stellarators, Reversed Field Pinches) or geometrical configuration (toroidal or linear devices) [1,2]: an evidence of a common physical mechanism driving this turbulence.

The experimental and theoretical investigation is rapidly progressing in the study of the properties of the individual coherent structures (see, e.g., [3]). However, statistical analysis of the signal is essential in order to gain information about the underlying physics. We carried out such a task earlier using data from the RFX experiment [4]: in a first paper [5], we performed an analysis of the Probability Distribution Function (PDF) for saturation current ($I_{sat}$), $P(I_{sat})$, measured by Langmuir probes at the plasma edge. But for a rather weak temperature dependence, $I_{sat}$ is proportional to the quantity of ions collected by the probe, hence to plasma density. The presence of coherent contributions in the signal translates into PDFs that differ from the canonical Gaussian distribution. An empirical fit of data, supported to some extent by theoretical considerations, suggested a lognormal form for $P(I_{sat})$. Lognormal distributions are naturally encountered in multiplicative (fragmentation) processes. Graves *et al* [6] performed recently a radial scan in the Scrape-Off Layer (SOL) of TCV tokamak, finding only a marginal accordance with the lognormal PDF, while most of their data were compatible with other analytical forms for *P*: they were interpreted within a many-levels birth-death-immigration problem, already studied in [7,8] and solved in terms of a Gamma function, featuring an exponential tail. To further support the idea that fragmentation mechanisms must be relevant to this study, it should be mentioned that recent numerical simulations describe the coherent structures as originating from the stretching and tearing of streamers from the plasma close to the separatrix (see, e.g. [9-12]). In a second paper [13], we attempted to provide an account of experimental data through a phenomenological fragmentation model. The model used therein describes $P(I_{sat})$ as the result of a fragmentation process of structures over a finite number of sizes, between a maximum and a minimum scale. The model was found to work well, providing a rather close agreement with the used data set, better than that used in [5]. It provided therefore further evidence in favour of the idea that fragmentation processes are decisive for understanding edge turbulence. It predicted a tail of the PDF at large values that, again, is approximately exponential, not lognormal. Independently from any theoretical consideration but simply as a result of visual inspection of the data, an exponential tail for $P(I_{sat})$ had already been suggested by other authors, see, e.g., [14] (and also [2], for the possible connections with the Bramwell-Holdsworth-Pinton distribution, frequently encountered with highly correlated finite-size systems). Unambiguously fixing the shape of the PDF may provide some insight about the still partially unknown physics driving the fluctuations' dynamics.

In this work we present new data produced by the Gas Puff Imaging (GPI) diagnostics [15] installed on the recently rebuilt RFX-mod device [16,17]. These data combine the improved acquisition frequency of the diagnostics (10 MHz instead of 1 MHz of earlier Langmuir probes) with an average increased duration of pulses in the new device, to obtain a still unreached (for RFX) quality in terms of statistical noise of the PDF as well as of density range scanned. This should, in principle, allow us to make a PDF analysis that is free from the ambiguities related to statistical and background noise.

Since the starting of its operations in late 2004, RFX-mod has been run under different regimes, from the "free" regimes, where no particular control of the plasma column is applied, to several "controlled" regimes (CR), aimed essentially at better controlling the magnetic configuration at the edge and reducing plasma-wall interactions. These CR's have by now become the new standard way of operating RFX, since they allow outstanding improvements with respect to the older configuration for the same plasma parameters (current and density): roughly, a 50% increase of temperature and energy confinement time, and up to a threefold increase of pulse duration [18]. They often feature improved transport properties at the edge, therefore, there is the possibility to study the statistical properties of the density fluctuations, trying to relate them to measurable differences in the transport in the plasma.

The outline of the work is as follows: in the next section a brief introduction to GPI diagnostics installed in RFX-mod is given. Section 3 provides an account of several experimental RFX-mod operative scenarios. Section 4 presents the statistical study, in terms of an analysis of the Probability Distribution Function of the GPI signal, for a selection of discharges from each of the scenarios listed in the earlier section. The visual inspection of a rather large data set of PDFs brought us to guess some common tracts encountered in most of them: namely, that the experimental PDF, rather than with a single analytical function, could be better interpolated using a linear combination of two (or possibly more) such curves. A simple phenomenological model is then developed in Section 5, to provide an account of these empirical findings. Results of Section 4 are thus interpreted in the light of this model in Section 6. Finally, Section 7 contains a summary and some conclusions.

**2. The GPI diagnostics**

A sketch of the diagnostics is given in Fig. (1). The system consists of a gas-puffing nozzle and 32 coplanar optical channels to measure the local radiation from an area normal to the local magnetic field. The optical lines are 5 mm spaced and cover an area of about 70 mm in the toroidal direction and 40 mm in the radial one. The effective bandwidth of the electronics is 2 MHz and signals are sampled at 10 Msamples/s. In this work the signal from one of the chords of the central fan will be used.

The gas is puffed into the plasma at a rate between $10^{18}$ and $10^{19}$ atoms/s. A detailed modelling of the gas within the plasma and with the walls would be needed in order to precisely assess its impact over the local plasma conditions (density and temperature), however previous experiences with gas puffing in RFX and Alcator C-Mod suggest that this rate of influx is not perturbative to the plasma [19,20].

In a first phase of operations, the gas puffed was Hydrogen, that is the working gas in RFX, and the measured wavelength was the $H_\alpha$ line at 656 nm. However, this setting up caused concerns about the signal-to-noise ratio, since the GPI signal is line-integrated: although most of the emission comes from the region where the gas is puffed in, a background signal due to gas desorbed from the wall on the opposite face of the torus is also present. In proper operating conditions, the brightness of the local cloud in front of the GPI is estimated one order of magnitude more intense than the background emission due to the influx from the first wall. However, all of RFX discharges are plagued by a localized kink perturbation, usually locked at the wall [21]. When any region of the intense plasma-wall interaction caused by this perturbation enters the view field, the two brightnesses become comparable. In order to reduce this spurious contribution the optics has been focussed onto the puffed cloud. This does not eliminate the wall contribution, but averages its fluctuations over a large length scale, hence probably cancelling them.

At a second stage, the problem was removed altogether by puffing Helium rather than Hydrogen, and measuring the HeI line at 668 nm, which stands clearly out of the Hydrogen spectrum. However, the discharges analyzed in this work include both kinds of gas puffing.

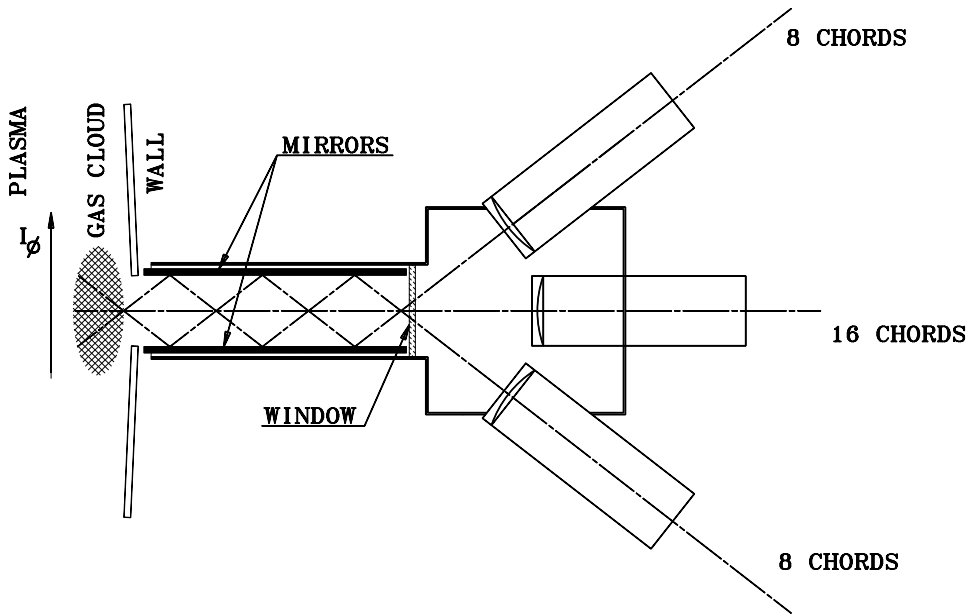

Fig. 1. Layout of GPI diagnostics. The instrument is aligned along the equatorial axis.

Since, to some extent, this work overlaps with affine studies carried on using data taken from Langmuir probes, it is useful to point out some issues concerning the data collected by the two diagnostics: I) GPI collects radiation mostly from atoms puffed into the edge plasma and excited by electron collisions. The collected signal is proportional to the collision frequency, hence to electron density, but depends also on temperature, that governs the electron-atom scattering rate. This dependence is different from the $T^{1/2}$ scaling that holds for the saturation current signal of Langmuir probes. Therefore, in principle, strong temperature fluctuations may lead to severe divergences between the two signals. (This is however not expected in RFX, where relative temperature fluctuations are smaller than density fluctuations- see [22,23]). II) It should also be pointed out that the active volumes for GPI and Langmuir probes are different, since the latter average over the flux tubes that cross them. Hence fluctuations might be averaged out somewhat differently.

A more important issue concerns the stationarity of the signal. The valves do puff gas at a slowly varying rate. The collected signal is proportional to the instantaneous quantity of gas available for excitation, and it must therefore be corrected for this slow variation before any manipulation. We proceeded as follows: the original time series was divided into 1 ms slices. Within such small time intervals, the variation in the gas inflow is negligible. The signal in each time slice was divided by its mean value. Provided that any plasma parameter also remained constant, the result of the division is just that of removing from the signal the influence of the source. After that, slices were reassembled into one single time series and analyzed. An example of the GPI signal before and after this manipulation is shown in Figs. (2a, 2b).

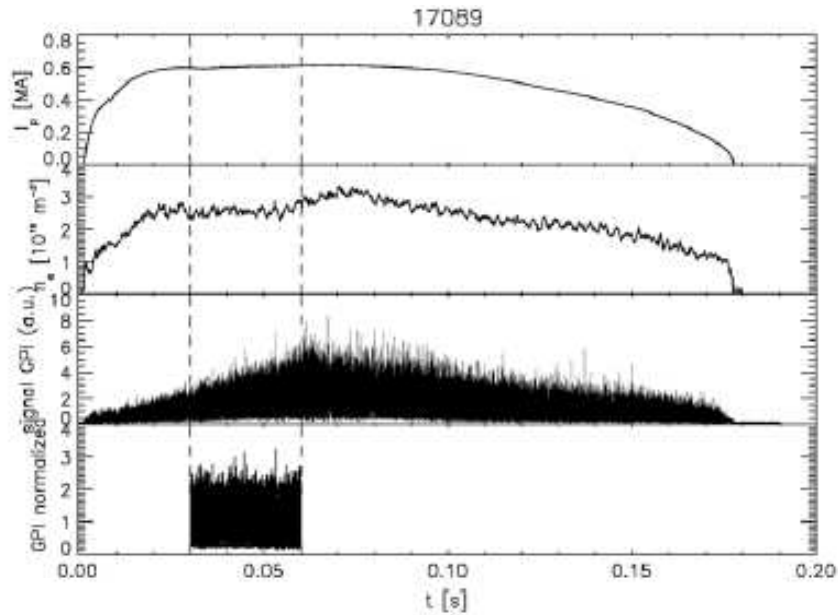

Fig. 2a. An example of discharge with main plasma parameters and GPI signal. Vertical dashed lines mark the temporal region used for statistical analysis. From top to bottom: plasma current; electron density; raw GPI signal (the drift of the source intensity is clearly visible); GPI signal after the correction for the source drift (for greater clarity shown only in the analysis region).

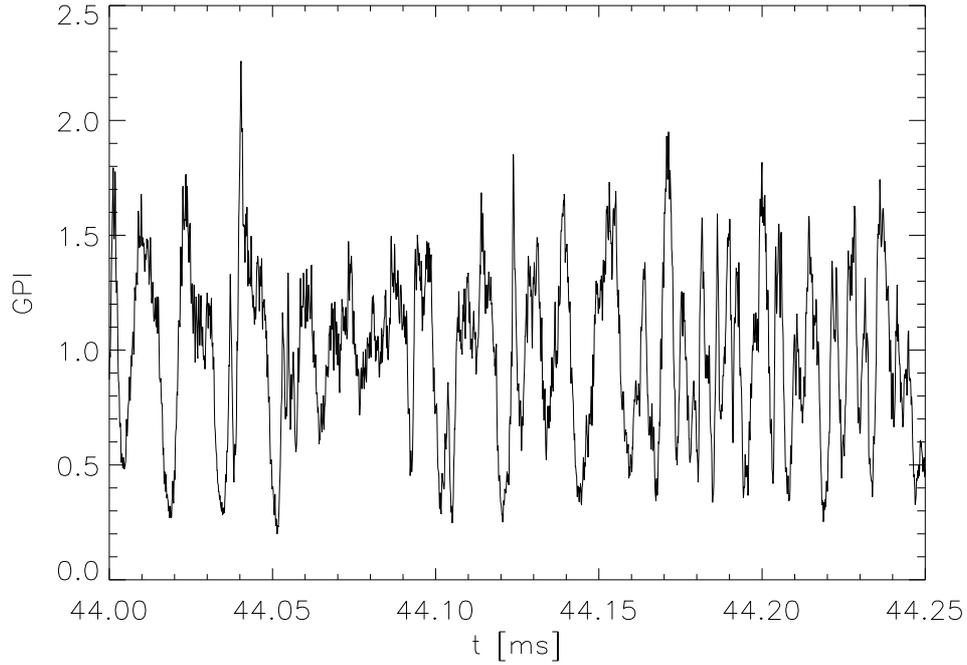

Fig. 2b. Zoom of the last panel of fig. (2a). Quite often, in experiments (see, e.g., [9,24-27]), the timestory of the signal is asymmetrical, showing first steeply rising fronts followed by smoother decay. Some examples of this same feature may be found in the figure, between 44 and 44.05 ms. This result is usually attributed to the asymmetrical shape of objects travelling in front of the measuring line of sight. Notice that in our case, due to the alignment of the lines of sight, the motion must be understood along the toroidal direction.

In connection with manipulations of the signal, it is worth recalling an aspect perhaps a bit overlooked in literature. In this kind of studies, it is the dynamics of the signal, *i.e.*, its fluctuations, that carries most of the relevant information, while the average part is ordinarily discarded as an irrelevant offset. We first commented in [5] that sometimes also the average part of the signal may carry useful information. We have just shown that, but for small corrections due to the temperature, the signal collected by either GPI or Langmuir probes is actually a quantity proportional to the plasma density, *n*. Since the density must be a positive definite quantity, discarding the average part, $<n>$, hides the constraint that the fluctuating part $\tilde{n} = n - <n>$ must be bounded on its negative side: $\tilde{n} \geq - <n>$. In order to avoid such problems, we will henceforth deal with the whole signal. The reader is therefore averted that, even when writing about fluctuations, we are actually referring to the non-zero-mean signal $<n> + \tilde{n}$, or better, given our normalizations, to $1 + \tilde{n}/<n>$.

## 3. An overview of RFX-mod operating regimes

Some words are in order about plasma control methods adopted in RFX-mod. The Reversed Field Pinch is a system in which the plasma produces its own equilibrium magnetic field configuration, characterized by a toroidal component of the magnetic field reversed at the edge with respect to the centre, through a dynamo mechanism [28]. The dynamo needs, to be sustained in a stationary state, some level of non-axisymmetric magnetic perturbations (although, ultimately, their origin is electrostatic [29]). Historically, the main limiting factor for RFP's performances has been the amplitude and the number of these perturbations. In order to reduce the unwanted aspects of dynamo modes, some plasma control techniques have been designed. Basically, all of these techniques involve shaping of the current/field profile through external coils.

*3.1 PPCD/OPCD*

Earlier methods are Pulsed Poloidal Current Drive (PPCD), or its variants like Oscillating Poloidal Current Drive (OPCD), where poloidal currents, applied to the plasma from the outside, transiently quench the need for dynamo action by the plasma, and hence reduce magnetic turbulence [30,31]. The poloidal currents at the edge affect the toroidal field, hence the reversal parameter $F = B_\phi(wall)/<B_\phi>$ is commonly adopted to monitor the driving (see Fig. 3).

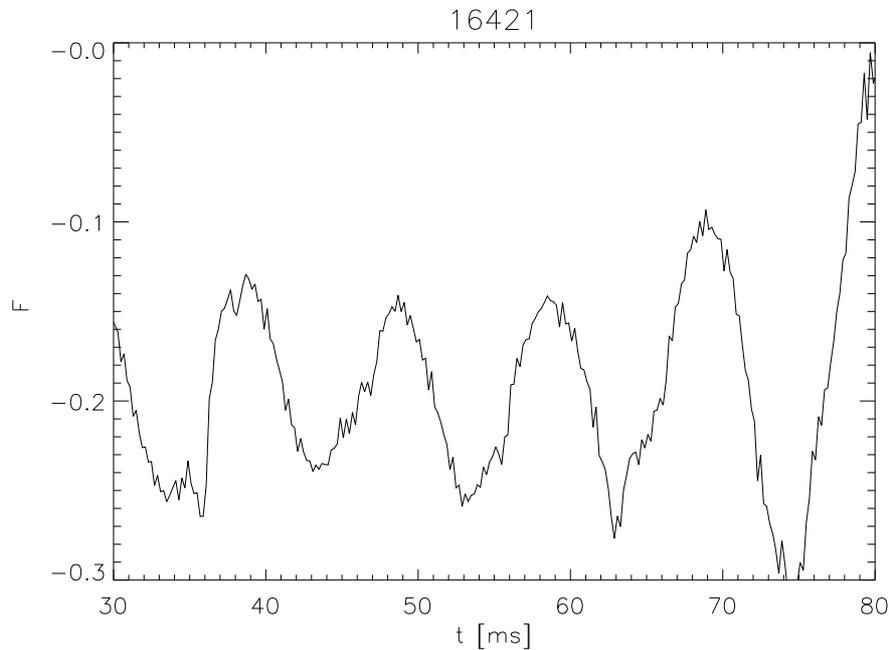

Fig. 3. Reversal parameter $F = B_\phi(wall)/<B_\phi>$ for an OPCD discharge.

*3.2 Virtual shell*

More recent techniques involve the sophisticated RFX-mod's "virtual shell" configuration, where external circuitry acts in feedback to cancel locally the radial magnetic field measured by a set of coils on the shell [16,17]. (Similar to this technique, is the cancellation of dynamo-driven edge currents from the outside in MST [32]).

*3.3 External rotating perturbation*

The coupling between dynamo modes and the shell tends to lock all the modes in phase between them and with respect to the laboratory frame of reference [21,33]. The resulting stationary localized perturbation is extremely deleterious in correspondence to those regions where the local shift of the magnetic field points outwards, thereby intercepting the wall. A way of mitigating the problem is to apply an *m*=0 (i.e., poloidally symmetrical), slowly rotating toroidally ($f \approx 50$ Hz) external perturbation that, by coupling with plasma modes, would drag them. This does not actually cure the problem of the plasma-wall interaction but spreads the outgoing flux over a much larger area, reducing the density of heat load. The dragging may not always be successful. In those cases, it is only the *m*=0 perturbation to rotate, modifying the structure of the edge turbulence, but to some degree still compatible with the monitoring by the GPI. Figure (4) displays the time traces of two relevant quantities during one discharge. The first is the contribution to the radial shift of the plasma column at the edge due to the rotating (*m*=0, *n*=1) mode, in the GPI measurement region. A negative slope means that locally the plasma is being shifted away from the wall, and the converse holds for a positive slope. Projecting the magnetic field line along the bent plasma column yields a radial component for the magnetic field; hence a non-zero slope may be understood also as an effective radial perturbation [34]. A cross-correlation analysis reveals the existence of a propagating perturbation in the GPI signal, moving toroidally with velocity $v_\phi$ (symbols in the plot). Although $v_\phi$ does not necessarily correspond to some sort of fluid flow, its value is consistent with an **E**×**B** drift, thus confirming earlier measurements done with Langmuir probes [35].

A complete analysis of the GPI signal in the presence of the external perturbation has been presented elsewhere [36,37]. It will be shown here that statistical properties of the fluctuations are affected by the phase of the external perturbation.

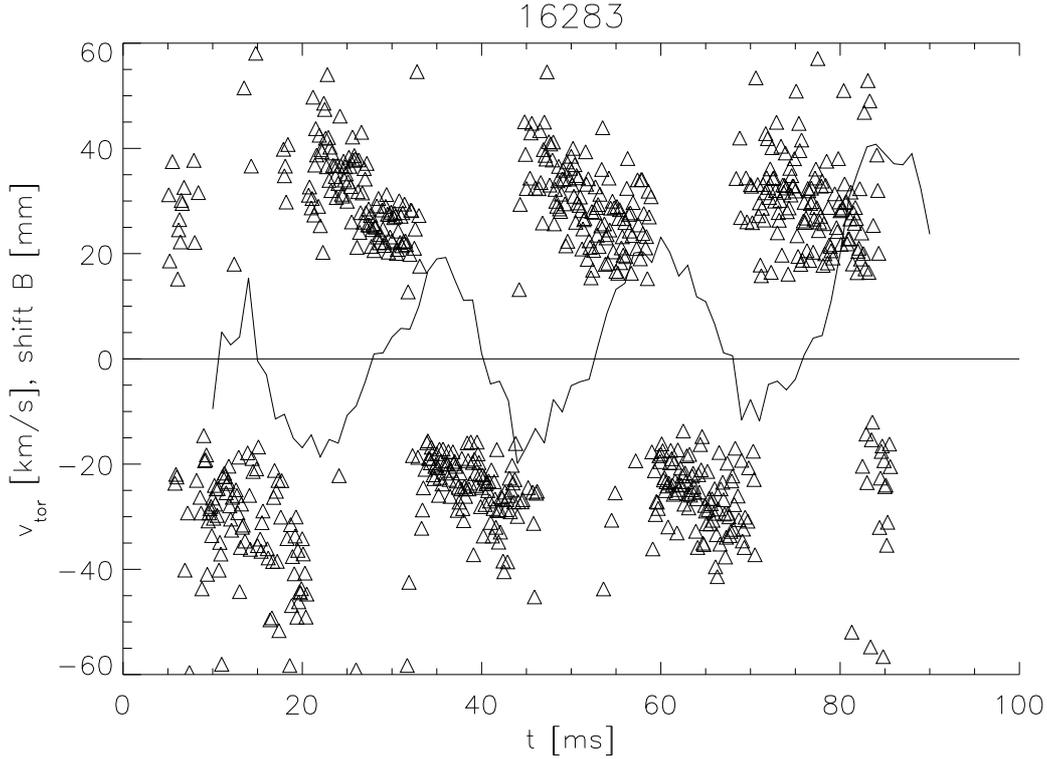

Fig. 4. Solid curve, radial shift due to the mode (*m*=0, *n*=1) at the toroidal position of the GPI. Triangles, toroidal velocity of the perturbation measured by the GPI.

*3.4 Quasi-Single-Helicity states*

As a consequence of some of the earlier control techniques-or even spontaneously-the plasma may change its properties and reach a configuration endowed with better confinement properties. The Single Helicity (SH) State is such a case. Such a configuration, where only one dynamo mode survives, is not turbulent, preserves good magnetic surfaces, and is expected theoretically to be a possible RFP stationary state, although until now has never been observed. The closest experimental evidence has been so far that of Quasi Single Helicity (QSH) states, where a single mode stands out from the others, that are, however, not negligible [38,39]. These states feature, too, remarkable improvements, e.g., in terms of temperature attained and fraction of preserved magnetic surfaces.

**4. Statistical analysis of GPI data for different RFX-mod operating regimes**

In the figures (5-8) we show PDF's extracted from GPI data for a sample of RFX-mod pulses for each of the operating conditions described above. Macroscopic parameters of the discharges were rather similar: maximum plasma current 0.6 MA, and core electron density about $3 \div 4 \times 10^{19}$ m$^{-3}$. For most pulses, except for those featuring mode rotation, $3\times10^5$ points (i.e., 30 ms) from the flattop quasi-stationary phase of the plasma current were binned. Notice that this figure is about one order of magnitude or larger than in earlier studies [5,13]. Pulses featuring external mode rotation were instead analyzed differently: the whole time series was sliced in intervals where the toroidal flow was either positive or negative, and each time region analyzed independently on the basis of the consideration that plasma properties might be considerably different. This way, for each pulse two or more subsets are obtained, lasting 5 to 10 ms each.

All pulses display PDFs whose tails, at large values of the signal, are approximately exponential. In some cases there are marginal features that could be attributed to noise or saturation of the signal. We attempted to fit the PDFs using Gamma functions, like done in [6]. The fit with just one Gamma function seems able to catch the shape of the whole PDF as far as uncontrolled pulses are concerned (Fig. 5), although some exceptions may be found. Improved Confinement pulses are instead less amenable to a single-function fit (see Figs. 6-8). However, when we attempted an empirical generalization with a linear combination of *two* Gamma functions, fits turned out to be fairly satisfactory:

$$P(n) = C_< \frac{(\beta_< N_<)^{N_<}}{\Gamma(N_<)} n^{N_< -1} \exp(-\beta_< N_< n) + C_> \frac{(\beta_> N_>)^{N_>}}{\Gamma(N_>)} n^{N_> -1} \exp(-\beta_> N_> n) \qquad (1)$$

where $n$ is the GPI signal, $\Gamma(N)$ is the mathematical function Gamma, $\Gamma(z) = \int_0^\infty t^{z-1} \exp(-t) dt$, introduced for normalization purposes, $C_<$, $C_>$ are positive numbers, with $C_< + C_> = 1$. The lowerscripts "<", ">" have been chosen to stress the order relation between the two parameters $N$: the "<" symbol labels the triple ($N$, $\beta$, $C$) with the lower $N$. We shall see later how the $N$'s may be attached to a physical meaning. In Table I we provide the fitting parameters ($N$, $\beta$) for the two fitting functions, as well as the relative weight of each Gamma functions, for all the curves plotted in figs. (5-8).

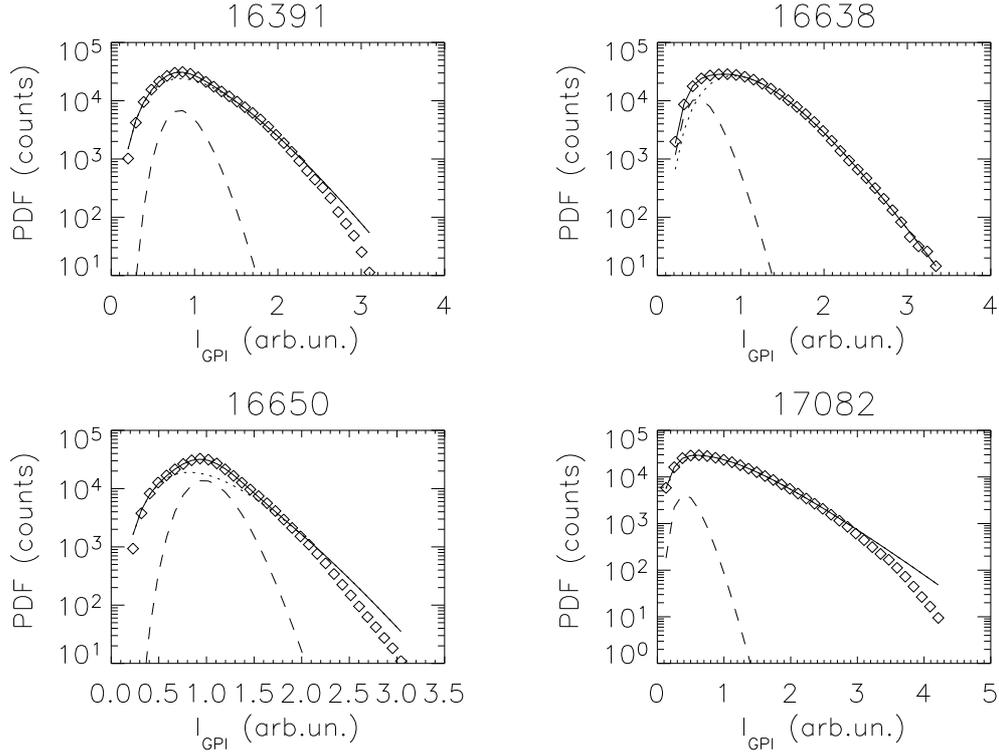

Figure 5. Examples of PDFs of $H_\alpha$ signal for uncontrolled RFX-mod discharges. Symbols, GPI data; solid curve, best fit from a linear combination of two Gamma functions (Eq. 1). The two individual components are shown as dashed and dotted curves (In pulse 17082, the dotted and the solid lines practically overlap). The fitting parameters, for this and the following plots, are summarized in Table I.

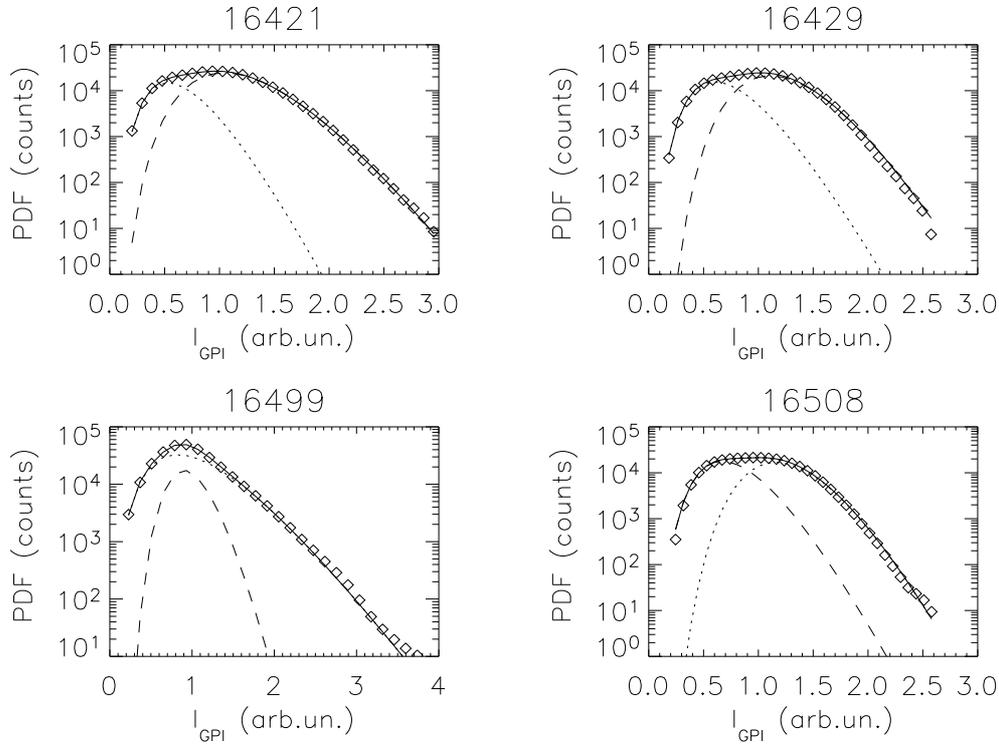

Fig. 6. Samples of $H_\alpha$ signal in improved confinement regimes (symbols). Upper row, OPCD pulses. Lower row, QSH pulses. Symbols, GPI data; solid curve, best fit from a linear combination of two Gamma functions (Eq. 1). The two individual components are shown as dashed and dotted curves.

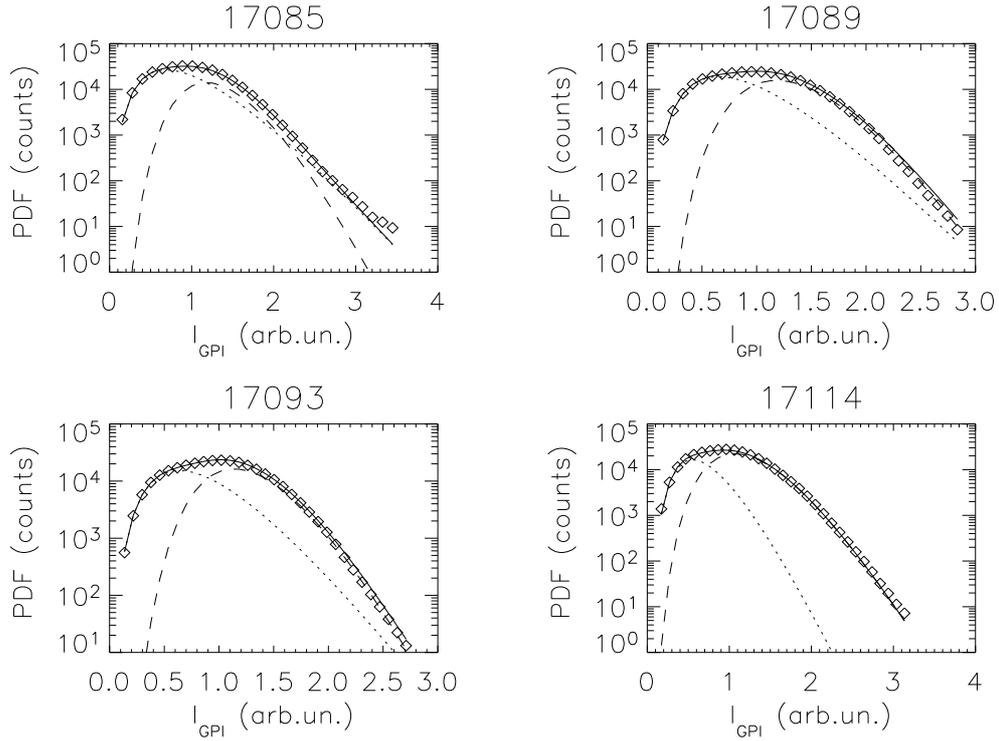

Fig. 7. Signals in the "virtual shell" configuration, measuring here the Helium line. Symbols, GPI data; solid curve, best fit from a linear combination of two Gamma functions (Eq. 1). The two individual components are shown as dashed and dotted curves.

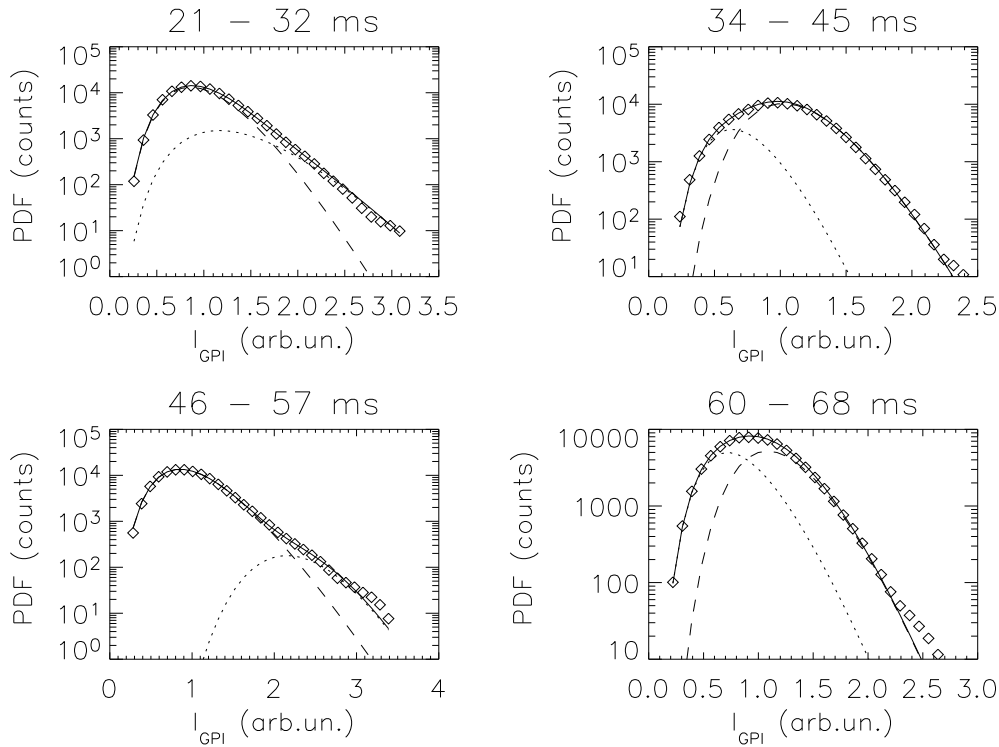

Fig. 8. PDFs from pulse 16283 featuring rotation of an external perturbation. Four time intervals are shown in correspondence to the switching of local plasma conditions (see Fig. 4). Symbols, GPI data; solid curve, best fit from a linear combination of two Gamma functions (Eq. 1). The two individual components are shown as dashed and dotted curves.

| Shot | $t_{in}$ (ms) | $t_{end}$ (ms) | $N_<$ | $N_>$ | $\beta_<$ | $\beta_>$ | $\%_<$ | $\%_>$ | discharge |
|---|---|---|---|---|---|---|---|---|---|
| 16391 | 30 | 60 | 5.3 | 18.6 | 0.98 | 1.15 | 88 | 12 | FD |
| 16638 | 30 | 60 | 6.5 | 9.8 | 0.93 | 1.85 | 86 | 14 | FD |
| 16650 | 30 | 60 | 5.5 | 21.1 | 1.00 | 0.98 | 70 | 30 | FD |
| 17082 | 30 | 60 | 2.9 | 7.9 | 0.96 | 2.16 | 96 | 4 | FD |
| 16421 | 30 | 60 | 8.1 | 11.2 | 1.65 | 0.87 | 27 | 73 | OPCD |
| 16429 | 30 | 60 | 8.7 | 15.6 | 1.48 | 0.85 | 35 | 65 | OPCD |
| 16499 | 30 | 60 | 5.3 | 20.2 | 0.99 | 1.06 | 78 | 22 | QSH |
| 16508 | 50 | 80 | 10.2 | 19.6 | 1.34 | 0.82 | 46 | 54 | QSH |
| 17085 | 30 | 60 | 4.7 | 14.7 | 1.14 | 0.79 | 67 | 33 | VS |
| 17089 | 30 | 60 | 5.4 | 15.1 | 1.25 | 0.79 | 56 | 44 | VS |
| 17093 | 30 | 60 | 5.3 | 15.2 | 1.28 | 0.82 | 50 | 50 | VS |
| 17114 | 30 | 60 | 6.7 | 10.8 | 1.55 | 0.85 | 32 | 68 | VS |
| 16283 | 21 | 32 | 8.6 | 9.9 | 0.76 | 1.05 | 14 | 86 | RM |
| 16283 | 34 | 45 | 12.0 | 16.5 | 1.45 | 0.92 | 22 | 78 | RM |
| 16283 | 46 | 57 | 7.7 | 32.0 | 1.02 | 0.45 | 98 | 2 | RM |
| 16283 | 60 | 68 | 9.1 | 14.5 | 1.24 | 0.86 | 46 | 54 | RM |
| 17464 | 30 | 70 | 10.5 | 13.2 | 1.72 | 0.92 | 18 | 82 | VS |
| 17747 | 30 | 50 | 14.0 | 16.6 | 1.28 | 0.88 | 39 | 61 | VS+QSH |

Table I. This table summarizes the relevant parameters for a sample of the shots analyzed, including all the shots shown in the figures above: the time range of analysis ($t_{in}$ and $t_{end}$); the effective degrees of freedom for the two Gamma components ($N_<$, $N_>$); the corresponding average value of the two PDFs ($\beta_<$, $\beta_>$, with $<n> = 1/\beta$); the fraction of the total PDF that is accounted for by each of the two components ($\%_<$, $\%_>$); a label tagging the discharges (FD = free discharge; OPCD = Oscillating Pulse Current Drive; QSH = Quasi-Single-Helicity; RM = Rotating Mode).

An inspection of the literature shows that similar results are not confined to RFX: we mention work from Tore Supra, whose data have been interpreted in terms of a Gaussian contribution plus an exponential (or probably Gamma) one (fig. 1 of ref. [40]), and the PISCES device, although not a fusion device, where clear traces of a similar double structure were found in *P(n)* [41,42]. Perhaps the clearest examples of a double structure come from TORPEX plasmas [3]. Furthermore, in several devices there is evidence that the shape of these PDFs varies, acquiring more and more of a Gaussian character while going from the wall to the plasma (perhaps crossing the source region of the coherent structures) [6,27,41,43-45]. Notice that Gaussian curves, strictly speaking, can hardly be accomodated in this context since their infinite support is incompatible with the constraint $n > 0$. However, Gamma functions, for $N \rightarrow \infty$, become practically indistinguishable from Gaussians. We guess therefore that there is a substantial amount of evidence in the literature in favour of the hypothesis that experimental PDFs could be fitted by the sum of *two* contributions. Empirically, it is here found that these two contributions may be quantified reasonably well using Gamma distributions with different weights and parameters. The two functions should stand, intuitively, for the contribution from two independent subsystems. In the next section we attempt to provide a rationale for the emergence of such kind of functions, and also suggest further possible refinements to this picture.

## 5. An analytical model

To provide an attempt to a theoretical understanding of earlier results, we postulate that the instantaneous measured signal, *n(t)*, be the average of *N* independent contributions:

$$n = \frac{1}{N}\sum_{i=1}^{N} x_i \qquad (2)$$

The meaning of $x_i$'s is that of effective degrees of freedom of the system. One can give a visual interpretation in terms of eddies overlapping in correspondence to the volume effectively scanned by the diagnostics, producing the recorded output. This makes an ansatz like (2) rather plausible. In the model each $x_i$ is thought of as a stochastic variable, endowed with its statistical distribution $f_i(x_i)$. We will suppose that the distribution is the same for each variable: $f_i = f \; \forall i$: in other words the $x_i$'s are independent identically distributed stochastic variables. Moreover, the $x_i$'s assume only positive values, legacy of the fact that the density cannot be a negative quantity. The number of contributions, *N*, is not fixed but left as a fitting parameter of the model, with the obvious constraint $N > 0$.

Calculations follow the standard characteristics' method for the demonstration of the central Limit Theorem (see, e.g., ref. [46]). We first introduce the Laplace transform of *P(n)*, $\hat{P}$:

$$\hat{P}(s) = \int_0^{\infty} \exp(-sn) P(n)\, dn \qquad (3)$$

It may be identified, at the same time, with the average value over *P* of the function exp(-*s n*): $\hat{P}(s) = \langle \exp(-s n) \rangle$. By combining this relationship with Eqns. (2,3) and with the independence of *x*'s, we may thus write

$$\begin{aligned}
\exp(-sn) &= \exp\left(-s\frac{1}{N}\sum_{i=1}^{N} x_i\right) = \prod_{i=1}^{N} \exp\left(-\frac{s}{N} x_i\right) \\
&\rightarrow \langle \exp(-sn) \rangle = \int \prod_{i=1}^{N} dx_i\, f(x_i) \exp\left(-\frac{s}{N} x_i\right) \qquad (4) \\
&= \left\{\int_0^{\infty} \exp\left(-\frac{s}{N} x\right) f(x)\, dx \right\}^N \equiv \left(\hat{f}\left(\frac{s}{N}\right)\right)^N = \hat{P}(s)
\end{aligned}$$

Hence, the PDF $P(n)$ is obtained through the inverse Laplace transform of Eq. (3). The exact form of $\hat{P}$ depends upon $f$. In the limit of infinitely many variables, we may expand $ln(\hat{f})$ in powers of $s/N$:

$$N \ln(\hat{f}) \approx N \ln\left\{\int_0^\infty (1 - \frac{s}{N}x + ...) f(x) dx\right\} = N \ln\left\{1 - \frac{s}{N}\langle x \rangle + ...\right\} \approx -s\langle x \rangle \qquad (5)$$

Hence, by replacing this result in (4),

$$P(n) = \delta(n - \langle x \rangle) \qquad (6)$$

That is, regardless of the analytical form of $f$, the PDF for the sum of infinitely positive variables collapses to the Dirac delta. Any finite support for $P(n)$ implies necessarily a finite value for $N$, and in this case we need a guess for $f$. This may be worked out in several ways. An appealing interpretation is in terms of pure thermodynamics: each degree of freedom plays the role of a thermodynamical system weakly in contact with a reservoir (the background plasma and possibly other degrees of freedom), with which it may exchange "energy" (density, in this case). The average value of each $x$ is trivially fixed: it is easy to recover from Eq. (2) that $<x> = <n>$, and the latter is a control parameter, in principle imposed from the outside, not related to fluctuations; therefore $<x>$ is a given constraint. This corresponds exactly to the canonical ensemble description, whose equilibrium distribution is the Boltzmann distribution, with $\beta = <n>^{-1}$:

$$f(x) = \beta \exp(-\beta x) \qquad (7)$$

Indeed, the Maximum Entropy Principle [47] would identify (7) from the outset as the most probable distribution of any system, provided that the only information available about it is just the average value $<x>$. It is interesting to notice that one could infer (7) also by assuming the validity of the model proposed in the earlier work [6-8] and exploiting the relationships between the two models. We shall briefly discuss this route in the Appendix. Regardless of the path followed to reach (7), once we insert it into (4), we get

$$P(n) = \frac{(\beta N)^N}{\Gamma(N)} n^{N-1} \exp(-\beta N n) \qquad (8)$$

where $\Gamma(N)$ is the already introduced mathematical function Gamma. This concludes the derivation. Intuitively, from Eq. (2), $N$ is an integer number. Since the difference between constrained ($N$'s integers) and unconstrained ($N$'s reals) minimization turned out not to be wide after a few tests, we preferred to maintain numerical simplicity and let the *N*'s vary over real numbers. Eq. (8) alone is not sufficient to account for all the features retrieved from GPI data and displayed in the figures above. Some more general expressions must be found. A straightforward generalization is obtained by considering a linear combination of two (or more) Gamma functions on the basis of the additive law of composition of probability, *P(x) = P_A(x) + P_B(x)*, provided that the two events A, B, be mutually exclusive. That is, there must be separate sources of the kind of fluctuations parameterized by Eq. (8), that eventually lead to Eq. (1).

This is the path that led to the fits shown in Figs. (5-8), first guessed on a purely empirical basis that eventually turned out to be quite reasonable. However, Eq. (1) can still be improved: so far, it has been postulated that $P_A$, $P_B$ have the same functional dependence: that is, that they are due to two distinct sources but ultimately driven by the same mechanism. What if the driving mechanisms are different? An appealing feature of the Maximum Entropy Principle is that it automatically generates distributions *f* corresponding to different sets of constraints: in the equations above we just used the information that we supposed to have about the first moment of the distribution, <*x*>. However, it is perfectly legitimate to ask what changes if supplementary information is available. The next, obvious step is to fix the second moment: $\sigma^2 = <(x-<x>)^2>$. This looks, furthermore, a more intuitive choice if one wishes to handle a transport of diffusive origin, since the stochasticity built into this kind of transport is naturally quantified in terms of the variance around a mean value. Within the Bayesian approach, this means that one knows in advance the value of $\sigma^2$, and uses this information to constrain the shape of the PDF.

The Maximum Entropy Principle provides in this case a Gaussian shape for *f*:

$$f \propto \exp\left(-\frac{1}{2}\frac{(x-<x>)^2}{\sigma^2}\right) \qquad (9)$$

Inserted into (4), it yields the following expression, to be Laplace-inverted:

$$\hat{P}(s) \approx \exp\left[\frac{s}{2}\left(\frac{s}{N}\sigma^2 - 2<x>\right)\right]\left\{\frac{\left|\frac{s}{N}\sigma^2 - <x>\right|}{\frac{s}{N}\sigma^2 - <x>} - \mathrm{Erf}\left[\frac{s}{N}\frac{\sigma}{\sqrt{2}} - \frac{<x>}{\sigma\sqrt{2}}\right]\right\}^N \qquad (10)$$

This expression is rather cumbersome and its inverse Laplace transform cannot be performed analytically for any value of the argument $n$. However, it is possible to show that the two limits do hold:

$$P(n) \approx \begin{cases} n^{N-1}, & n \to 0 \\ \exp\left[-\dfrac{(n-<n>)^2}{2\sigma^2/N}\right], & n \approx O(<n>) \end{cases} \qquad (11)$$

Thus, it is possible to write down an analytical approximation for $P(n)$ by taking the product of the two limits (11):

$$P(n) \approx n^{N-1} \exp\left(-\dfrac{1}{2}\dfrac{(n-<n>)^2}{\sigma^2/N}\right) \qquad (12)$$

Fig. (9) shows a sample of $P(n)$ evaluated from numerical inversion of Eq. (10)[1], together with its approximation (12). Furthermore, a comparison with a Gamma distribution (8) is provided.

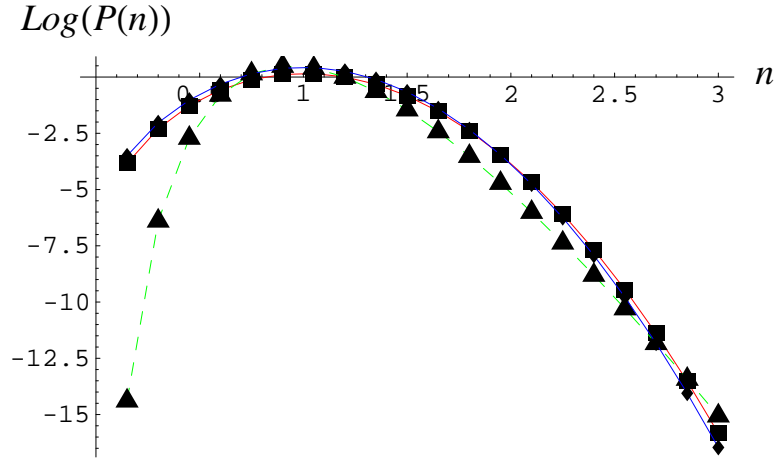

Fig. 9. Green dashed line with triangles, Gamma distribution from Eq. (8) with $\beta^{-1} = <x> = 1$, $N = 16$. Red solid line with triangles, $P(n)$ evaluated from Eq. (10) with $\sigma = \frac{1}{2}$, $<x> = 1$, $N = 2$. Blue solid line with squares, $P(n)$ from Eq. (12). We stress the huge difference in $N$ parameters needed to roughly match the tails of the two kinds of curves.

## 6. Discussion of the empirical fittings in terms of the model

The results presented in the previous sections are intriguing from a conceptual side, since they hint to a possible connection between the transport mechanisms of energy and matter at the edge and the statistical distributions: a rather common feature is that one of the two contributions has a small value for $N$ ($N_<$ : say, $N_< < 10$), while the second a large value ($N_>$: $N_> > 10$). An intriguing hypothesis, on the basis also of literature [6,40,41], is postulating the $N_>$ contribution as due to the uncorrelated fluctuations that drive turbulent diffusion, while the $N_<$ one is related to the coherent part of the fluctuations. Convection-driven flux, being the result of correlated movement of many particles, may be quite naturally related to this latter contribution. In other words, we might speculate that convective transport drives at the edge fluctuations whose statistical distribution is given by Eq. (8). Diffusive transport, instead should lead either to a similar PDF, but with a much larger value for the parameter *N* (practically infinite), or to a PDF like that given in Eqns. (10,12): from the practical point of view, it is impossible to discern between these two latter options: in the low-*n* region, the gaussian contribution to the curve (12) turns out to be order unity, hence the two functions (8) and (12) have, to leading order, the same power-law trend in the low-*n* region. Where the two curves do really differ is in the high-*n* wing: there, no matching is possible over large enough intervals, since the former curve is an exponential and the latter a Gaussian. In Figs. (5-8) the high-*n* tail looks, rather uncontroversially, an exponential, hence a PDF of the kind (12) may, at best, account for the low-*n* region. Indeed, a few tests were performed, by fitting a few of the PDFs from the Section 4 using a linear combination of Eq. (8) and Eq. (12), but no appreciable differences with the results already summarized could be found. In the following we discuss the results (5-8) in terms of the fit provided by Eq. (1). .

---

[1] The numerical inverse Laplace transform was done using the Mathematica® package *NumericalLaplaceInversion*, written by P.P.Valkó and J. Abate, and available at the URL: http://library.wolfram.com/infocenter/MathSource/4738/ .

We stress again that the identifications between low-*N*/high-*N* PDFs with coherent/incoherent structures and then with convective/diffusive transport must be understood as speculative, until now not really supported by definite proofs. Nonetheless, we claim that the results presented above provide rather strong evidence favouring this hypothesis. We add also two further comments to this regard.

The first evidence is that improved confinement regimes appear often related, at the edge, to a reduction of the content of the low-*N* fluctuations (see Table I). Conversely, the high-*N* contribution, that in free conditions has a small weight, gains relative importance under these operating conditions. Extensive research on the edge transport of RFX was carried on in the past, and showed that intermittent events, related to coherent structures, may account for up to 50% of edge flux. It is likely, therefore, that improvements in the edge transport may be related to reduction of the content of these structures.

Indeed, a very recent study strengthens the idea that the $N_<$, $N_>$ contributions are related to the coherent and incoherent part of the transport: Farge, Schneider and Devyinck [48] applied the Orthogonal Wavelet Method (OWM) to the analysis of ion saturation current signal on Tore Supra Tokamak. These authors assert that the OWM is able to disentangle the raw signal into its coherent and incoherent part. The shape of the latter contribution must be guessed in advance, and in [48] the reasonable ansatz of a Gaussian was done, after which the coherent contribution turned out to be quite close to a Gamma distribution. The overall result is fairly in accordance with our two-curves fits, although the OWM cannot deal fully with the more complicated two-Gamma fits. The next issue to discuss is therefore the existence of a wide range of variation of the parameters of the two components over the total PDF, as evidenced by Table I.

The fitting curve (1) involves six free parameters. The normalization of the PDF and the constraint $<x> = 1$ imposed on the signal allow removing two of them:

$$C_1 + C_2 = 1, \quad \frac{C_1}{\beta_1} + \frac{C_2}{\beta_2} = 1$$

With four parameters left and with a database amounting to few tens of discharges, any attempt to identifying any clear relationships between the parameters and discharge conditions would still be a demanding task. We believe, however, that some trends may be identified:

1) In free discharges (Fig. 5) the bulk of the PDF is quasi-Gaussian, while the tail is strongly skewed.

2) In discharges with some kind of mode control (OPCD, Virtual Shell, etc …, Figs. 6,7), on the contrary, the quasi-Gaussian component rules the higher-than-average part of the fluctuations.

Earlier in the text we mentioned that the literature is unanimous in relating changes of the shape of the PDF to the distance from the wall, hence to mutating background plasma conditions [6,27,41,43-45]. Indeed, this kind of phenomenology is consistent with a picture in which structures are born from within a radially extended region and then move mainly outwards: the more the probe moves into the plasma, the lesser the fraction of structures that intercept it.

In tokamaks, measurements are normally performed in the Scrape-Off-Layer (SOL) or just inside the separatrix. RFX has no physical limiter but the deformation of the plasma column creates an effective Scrape-Off Layer. The resulting SOL depth, $\lambda_{SOL}$, is not constant but depends on the local magnetic deformation. It should be kept in mind, furthermore, that, because of the relatively high level of magnetic fluctuations, in RFPs magnetic surfaces cannot be defined as precisely as in tokamaks. Accordingly, throughout this text $\lambda_{SOL}$ must not be understood as a clear-cut value, rather as (something better than) an order-of-magnitude estimate. Its average value over the whole surface of the torus, $<\lambda_{SOL}>$, must increase with the amount of the deformation. Qualitatively speaking, we may therefore envisage the following situation:

1) Free discharges feature on the average a worse control of the plasma column, hence a larger $<\lambda_{SOL}>$. The GPI diagnostics, which looks at fixed position close to the wall, in these cases is monitoring a region of "far SOL".

2) In controlled discharges $<\lambda_{SOL}>$ is, on average, smaller, hence the plasma conditions probed by the GPI are closer to those of near SOL.

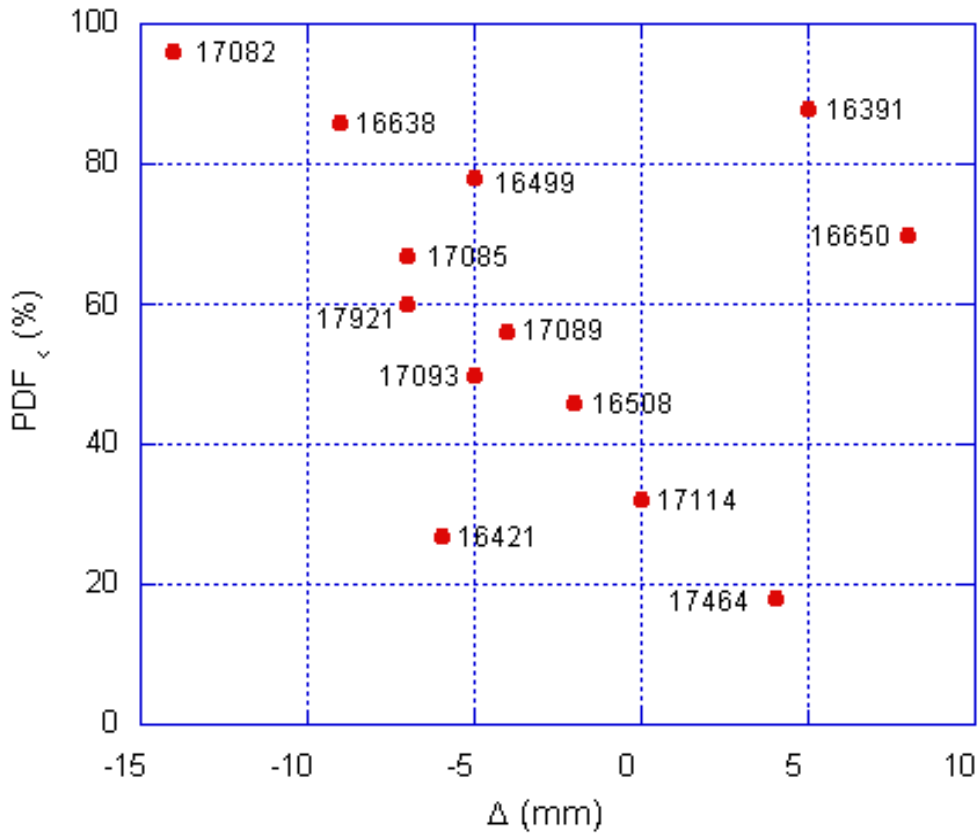

Fig. 10. It is shown the fraction of the PDF due to the low-*N* Gamma contribution (for reference, look at Table I) for a few discharges *versus* the shift of the plasma column at the GPI location.

In order to make visual these statements, we plot in Fig. 10 the fraction of the PDF due to the low-$N$ Gamma function ("the coherent contribution") *versus* the local displacement of the plasma column at the GPI measurement position, $\Delta$. For practical purposes, we measure $\Delta$ as the algebraic sum of the shifts due to $m=0$ and $m=1$ modes, as obtained from edge magnetic field probes. We may roughly assume that $\lambda_{SOL} \approx -\Delta$ for negative $\Delta$'s. The meaning of a positive $\Delta$ is that of a local displacement of the plasma beyond the wall, due to the tail of the locked perturbation. In this situation, the concept of SOL loses its meaning. Starting from large negative values for $\Delta$ and going toward zero is equivalent to sampling SOL regions closer and closer to the separatrix, and we find accordingly a larger and larger quasi-gaussian contribution (i.e., reduced coherent fraction). There is also a turning point, around 0 mm, beyond which the quasi-gaussian contribution inverts its trend. Actually, the GPI is no longer probing the SOL but is fully embedded into the main plasma, whose conditions and ruling physics are severely different.

It is interesting to notice that discharges with Virtual Shell active and simultaneously featuring QSH states display often "effective" quasi-gaussian PDFs: both Gamma contributions have large (> 10) values for $N$ (and often the two values $N_<$, $N_>$ are numerically close to each other). This seems suggesting that assigning the $N_<$ contribution to some "coherent" mechanism does not appear adequate in this case. Hence, we have not included those pulses in the figure, but for a "marginal" case (pulse 16464 that, actually, lies at the very bottom in fig. 10). The result appears rather reasonable by considering that these discharges should implement the best controlled magnetic configurations.

The case featuring the rotating perturbation in principle may provide further information, since the plasma column oscillates to and fro within one and the same discharge between a minimum and a maximum displacement. The results may be extracted from Table I. The relationship between the shift $\Delta$ and the relative percentage of coherent/incoherent contributions becomes, of course, meaningless now. The place of the parameter $\Delta$ seems now taken by its time derivative: $d\Delta/dt$ or, equivalently by the sign of the radial component of the magnetic field, which is proportional to the former quantity, $b_r \propto d\Delta/dt$. When $b_r$ is pointing towards the wall ($b_r > 0$) the coherent fraction becomes dominant, and, conversely, when $b_r$ changes sign it is the quasi-gaussian part to dominate. Here, too, just like the (VS+QSH) case cited earlier, Table I must be read properly in order not to misunderstand these statements. Actually, during the (21-32 ms) positive $b_r$ phase both contributions may be classified as quasi-coherent (both have low $N$'s < 10). Similarly, during the (34-45 ms) negative $b_r$ phase, both contributions are quasi-gaussian ones ($N > 10$). The explanation for this behaviour is less clear than in the static case, and we do not advance any suggestion.

## 7. Concluding remarks

In this work we have presented an extensive review of some statistical features of edge turbulence under several different operating conditions. The results presented here allow concluding that, as far as edge turbulence is concerned:

(I) the phenomenology encountered in RFX-mod is not unequivocal, apparently depending to a large extent on the confinement regime. It shares however common features with other devices, thereby supporting the idea of the universality of turbulence at the edge.

(II) We have been able to cast most of the phenomenology into a rather simple picture based upon few parameters. An intuitive *description* for its statistical features is provided in terms of an analytical model. An *interpretation* of the phenomenology is also attempted in terms of convective-diffusive related transport. It appears consistent with evidences from several other experiments [6,27,41,43-45], and is also supported by independent theoretical analyses [48].

III) Assessing that the two-Gamma-functions picture is adherent to reality and not just a useful but artificial mathematical parameterization, and even providing a rationale for their insurgence is, of course, just a necessary first step. The next desirable step to do would be to relate the purely phenomenological parameters with detailed mechanisms. Within the picture provided in this work, this is tantamount to understanding what are the mechanisms that regulate the relative weights of the two contributions (this would correspond to giving a physical meaning to the values of the parameters $C$ in Eq. 1), the average density driven by each contribution (the parameters β), and the amplitude of the fluctuations around these averages (which, lastly, provide values for the $N$'s). We were able to partially pursue this task by recognizing the relevance of the distance of the confined plasma from the measurement region. This is consistent with common findings from literature, where the distance from the separatrix into the SOL was identified as a crucial parameter towards the characterization of the PDF.

It is important to recognize also the limitations inherent to the present approach. Indeed, we emphasized since the outset that the inspiring guideline behind this study has been to identify common tracts in turbulence phenomenology between different machines. However, a complete understanding of the problem must deal not only with similarities but also with differences that necessarily exist: any two devices (or classes of devices, such as tokamaks and RFPs) *are* different (often largely different) in terms of plasma density, temperature, magnetic field, … . An appropriate analogy may be drawn between the present approach and the derivation of the velocity distribution function for an ideal gas: there, considerations based on fairly general principles (symmetries and conservation laws) lead straightforwardly to writing down the Maxwellian PDF. At that stage, any two gases are identical, differing at most for their temperature. But real gases are not ideal: any further refinement necessarily involves some knowledge of the specific molecule-molecule interaction, thereby destroying the generality of the solution. The approach carried forth in this work, as well as in many of the references quoted, is clearly similar to identifying the plasma's "Maxwellian". Within this context, the parameter $N$ acts as the equivalent of the "temperature" of the system, while $\beta$ plays the role of a rigid velocity of translation. They are moments of the distribution function, which are fixed by boundary conditions and may provide only coarse-grained information about the microscopic behaviour.

**Acknowledgements**

T. Bolzonella, D. Escande, M. Farge, R. Paccagnella, D. Terranova, M. Valisa and P. Zanca are acknowledged for their help with various aspects of the manuscript. This work was supported by the European Communities under the contract of Association between Euratom/ENEA. The views and opinion expressed herein do not necessarily reflect those of European Commission.

**Appendix**

In this appendix we will show that it is possible to assign an analytical expression for $f$ (Eq. 7) using, under a suitable limit, the Jakeman's model implemented in [6-8]

We note first of all that, if $N = 1$, $P(n)$ and $f$ must stand for the same quantity, $P \equiv f$ (see Eq. 4). Hence, evaluating $P(n)$ under this limit, provides the sought expression for $f$.

We do not give here details of the Jakeman's model, for which we refer the reader to the papers [6-8]. We just notice that, there, the tendency to clustering (i.e., allocating all the population into few levels of the system) is there quantified by the parameter α: α = 1, ∞, are the two extremal cases of maximal and null clustering. Translated into the present model,, α = 1 must correspond to the case where all the signal is carried on by just one degree of freedom, hence to $N = 1$. However, when $\alpha = 1$, the Jakeman's result for $P$, $P_J$, collapses to a simple exponential. Making the ansatz that the two models must yield the same results, $P_J \equiv P$, implies that $f$, too, must be an exponential, and we get Eq. (7).

Although rather straightforward, this derivation is purely academical, since it must rely on the validity of an independent model which is already able to provide by itself the full answer (8) . There is also another unpleasant aspect in that one is forced to use the limit (α, $N$ ) → 1. Although mathematically legitimate, this limit conflicts with the requirement $P(n = 0) = 0$. In other words, only cases with $N > 1$ possess a physical sense.